\documentclass{appolb}
\usepackage{graphicx}

\begin{document}
\title{Search for polarization effects in the antiproton production process
}
\author{D. Grzonka, K. Kilian, J. Ritman, T. Sefzick
\address{Institut f\"ur Kernphysik, Forschungszentrum J\"ulich, 52425 J\"ulich, Germany}
\\
W. Oelert
\address{Johannes Gutenberg-Universit\"at Mainz, 55099 Mainz, Germany}\\
 M. Diermaier, E. Widmann, J. Zmeskal
\address{Stefan-Meyer-Institut f\"ur subatomare Physik, 1090 Wien, Austria}\\
 {B. Glowacz, P. Moskal, M. Zielinski}
\address{Institute of Physics, Jagiellonian University, PL-30 -059 Krakow, Poland}\\
 {M. Wolke}
\address{Department of Physics and Astronomy, Uppsala University, 75120 Uppsala, Sweden}\\
 {P. Nadel-Turonski}
 \address{Thomas Jefferson National Accelerator Facility, Newport News, Virginia 23606}\\
 {M. Carmignotto, T. Horn}
\address{Physics Department, The Catholic University of America, Washington, DC 20064}\\
 {H. Mkrtchyan,  A. Asaturyan, A. Mkrtchyan, V. Tadevosyan, S. Zhamkochyan}
\address{A. I. Alikhanyan Science Laboratory, Yerevan 0036, Armenia}\\
 {S. Malbrunot-Ettenauer}
\address{CERN, Physics department, CH-1211 Genève 23, Switzerland}\\
 {W. Eyrich, F. Hauenstein, A. Zink}
\address{Physikalisches Institut, Universit\"at Erlangen, 91058 Erlangen, Germany}\\
}
\maketitle
\begin{abstract}
For the production of a polarized antiproton beam various methods have been suggested including the possibility that antiprotons may be produced polarized which will be checked experimentally. 
The polarization of antiprotons produced under typical conditions for antiproton beam preparation will be measured at the CERN/PS. If the production process creates some polarization a polarized antiproton beam could be prepared by a rather simple modification of the antiproton beam facility.
The detection setup and the expected experimental conditions are described.  

\end{abstract}
\PACS{PACS numbers come here}
  
\section{Introduction}
Polarization observables  reveal more precise information of the structure of hadrons and their interaction, and the disentangling of various reaction mechanisms is often only possible by controlling the spin degrees of freedom.
With beam and target particles both being polarized quantum states can be selectively
populated. For example, in $\bar{p}-p$ reactions 
a parallel spin configuration of antiproton and proton ($\bar{p}\uparrow \ p\uparrow$)
is a pure spin triplet state and with an antiparallel spin configuration  
($\bar{p}\uparrow \ p\downarrow$) the spin singlet state is dominant.
The possibility of adjusting different spin configurations
is important for various topics in the regime of high as well as low energy.

While polarized proton beams and targets 
are routinely prepared the possibilities for the preparation 
of a polarized antiproton beam are still under discussion.
Proposals for the generation of a polarized antiproton beam have already been presented before the first cooled antiproton beams were available. 
Methods that have been discussed are: hyperon decay, spin filtering,
  spin flip processes,
 stochastic techniques,
  dynamic nuclear polarization,
  spontaneous synchrotron radiation,
  induced synchrotron radiation,
  interaction with polarized photons,
  Stern-Gerlach effect,
  channeling,
  polarization of trapped antiprotons,
   antihydrogen atoms, and also a possible
  polarization of produced antiprotons.
 Summaries of the various possibilities can be found in \cite{bod85}, \cite{ste08}, \cite{ste09}, \cite{mey08}.
Most of the methods are not usable due to the extremely low expected numbers 
of polarized antiprotons or the low degree of polarization
and for some methods reasonable calculations are not possible since relevant parameters are not known.
Due to the large required effort no feasibility studies have been performed so far.
 
A well known source for polarized antiprotons is the decay of  $\bar{\Lambda}$ into
$\bar{p} \, \pi^+$ with 
a $\bar{p}$ helicity of 64.2 ($\pm$ 1.3) \%  (the more precise value for $\Lambda$ decay is taken)
in the  $\bar{\Lambda}$ rest frame \cite{pdg10}.
By measuring the direction and momenta of the 
$\bar{\Lambda}$, the $\bar{p}$, and 
$\pi^+$ in the laboratory system,
the decay kinematics can be reconstructed and the
transversal and longitudinal antiproton polarization components 
in the lab system for each event can be determined.
The method was used at FERMILAB in the only experiment with polarized 
antiprotons so far \cite{bra96} 
studying the polarization dependence of inclusive $\pi ^{\pm} \pi ^0$ production.
A proton beam of 800 GeV/$c$ momentum 
produced antihyperons ($\bar{\Lambda}$) and their decay antiprotons with momenta 
around 200 GeV/$c$.
A mean polarization of 0.45 was observed
but the polarized antiprotons do not constitute a well defined beam and 
at lower energies the situation further deteriorates.
For the preparation of a pencil beam of polarized antiprotons other methods are required.

Presently the most popular proposal is the filter method 
on a stored antiproton beam, by which 
one spin component is depleted due to the spin dependent hadronic interaction
of a beam passing a polarized target.
The filter method in a storage ring 
was first suggested 1968 
in order to polarize high energy protons in the CERN ISR \cite{cso68}. 
For filtering of antiprotons at lower momenta it was later pointed out 
that the new technique of phase space cooling 
is mandatory and the relevant parameters were shown \cite{kil80}, \cite{kil82}.
In 1993 a feasibility study of the filter method with phase space cooling 
was performed with a proton beam on polarized protons at the TSR in Heidelberg
showing clearly the buildup of polarization \cite{rat93}.
A polarization of about 2\% was achieved after 90 minutes filtering time.
For antiprotons the filter method with cooling should also work
if one can find any filter interaction with both large spin-spin dependence and cross section.
 Till now there
are no data for the spin-spin dependence of the total $\bar{p}p$ cross section.
From theoretical predictions one expects that 
longitudinal polarization effects are larger than transversal effects \cite{ric82}, \cite{bar09},
\cite{dmi08}, \cite{dmi10}.
The consequence might be that a Siberian snake is needed in the filter synchrotron.
Experimental $\bar{p}p$ scattering data are needed to work out the conditions
and expected properties of a polarized $\bar{p}$ beam prepared by the filter method.
The PAX collaboration is working on this topic with polarized protons as filter  \cite{len05}, \cite{oel09}, \cite{bar09}. The spin filter method has been demonstrated at COSY for transverse polarized protons and similar studies 
for longitudinal polarization are planned \cite{aug12}.
Another filter reaction with large polarization effect,
which was recently proposed \cite{sch10} is the interaction with polarized photons but the low intensity of available photon beams together
with the small cross section makes it unlikely to produce sufficient polarized antiprotons for significant experiments.

A simple possibility for a polarized antiproton beam may be the production process itself.
If the antiproton production process creates some polarization it would be rather simple to prepare a polarized antiproton beam.

\section{Search for polarization in antiprotons production}
\label{search}
It is well known that particles, like e.g. $\Lambda$-hyperons, produced in collisions of high energy unpolarized protons show a significant degree of
polarization \cite{ram94}. 
Maybe also antiprotons are produced with some polarization but up to now 
no experimental studies have been performed in this direction.

\begin{figure}[hbt]
\includegraphics[width=0.65\textwidth, clip=true]{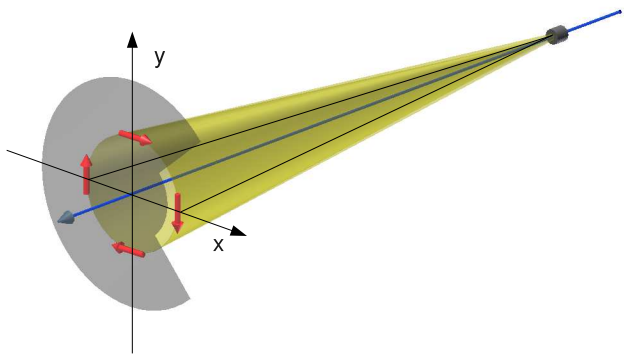} 
\begin{minipage}[b]{0.3\textwidth}
\caption{
Possible polarization vector 
for a given production polar angle $\theta$, indicated by the red arrows.
In order to select an antiproton beam with transverse polarization
one spin direction has to be separated by a suitably shaped
absorber plate as indicated in grey.
}
\label{spincut}

\end{minipage}

\end{figure}

The production of antiprotons 
is typically done by bombarding a solid target 
with high momentum protons. At CERN the beam momentum is about 24 GeV/$c$
and the number of collected antiprotons is in the order of one per $10^6$ beam protons.
The production mechanism seems to be a rather simple quasi-free $p$-nucleon interaction.
The $\bar{p}$ momentum spectrum which is peaked around 3.5 GeV/$c$ is consistent with a pure
phase space distribution  for a four particle final state:
$p p \rightarrow p p \bar{p} p$.
The basic process is a creation of baryon-antibaryon 
out of collisional energy.
 If transverse polarization occurs  
(Fig. \ref{spincut}) then a polarized 
beam can be prepared in a rather simple and cheap way by blocking
up and down events, and
one side of the angular distribution.
Furthermore the pure S wave region around 0 degree ($<50 \, mrad$) has to be removed.
A simple modification of the extraction beam line 
of an existing $\bar{p}$ production and cooler facility
with absorbers would be sufficient to extract a polarized beam. 
Of course one has
to avoid polarization loss in de-polarizing resonances in the
accumulator cooler synchrotron. This has to be taken into account when such a facility will be constructed.
But first of all it has to be investigated whether the production process creates some significant polarization.
  
\section{Asymmetry measurement of produced antiprotons}
In view of a future polarized antiproton beam the polarization studies will be performed at the typical
conditions for antiproton beam preparation.
Actually the only facility for antiproton beams is the AD at CERN where the antiprotons, produced with the 24 GeV/c PS proton beam, are injected at a momentum of 3.5 GeV/c. Similar conditions are foreseen for the future FAIR facility.

In order to measure the polarization of the produced antiprotons
a scattering process with known and sufficiently high analyzing power has to be performed.

\begin{figure}[hbt]
\begin{center}
\includegraphics[width=0.6\textwidth]{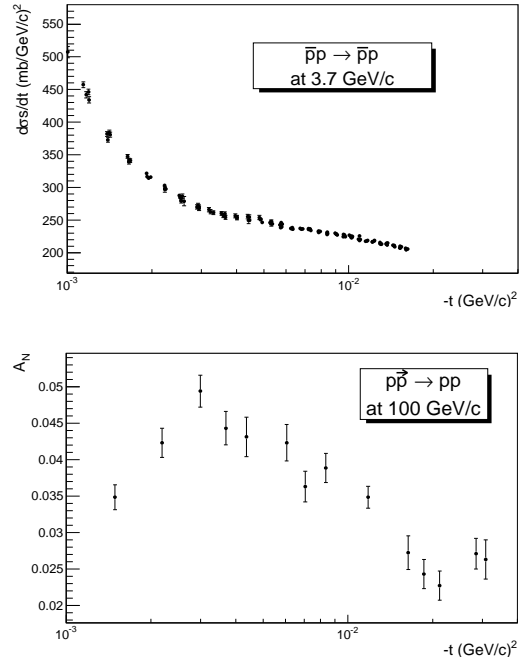} 
 \end{center}
\caption{Cross section data for elastic antiproton proton scattering at a beam momentum of 3.7 GeV/c \cite{arm96} (upper part) and analyzing power of proton-proton scattering \cite{oka06} (lower part) as a function of the four momentum transfer t. 
The data for the analyzing power have been taken with a proton beam momentum of 100 GeV/c at a polarized atomic hydrogen target. The analyzing power of antiproton scattering is  of the same size with opposite sign. 
\label{dsan}}
\end{figure}

Well known and calculable is the analyzing power in the high energy $p p$ elastic scattering 
in the Coulomb nuclear interference (CNI) region. The analyzing power in the CNI region 
at high energies is attributed to 
the inference between a non-spin-flip nuclear amplitude and an 
electromagnetic spin-flip amplitude \cite{kop74},\cite{but99}, \cite{gro90}, \cite{oka06}.
The maximum analyzing power is approximately given by \cite{kop74}, \cite{akc93}:
$A_{N}^{max} = \sqrt{3}/4 \cdot \sqrt{ t_p}/m \cdot (\mu-1)/2$, 
with $\mu$=magnetic moment, $m$=mass.
A maximum of 4-5\% is reached and
the four momentum transfer at the peak $t_p$ is given by:
$t_p = -8\pi \sqrt{3} \alpha/\sigma_{tot} $,  with the fine-structure constant $\alpha$ and the total cross section $\sigma_{tot}$, which results in a value of $t_p  \sim -3 \cdot 10^{-3} (GeV/c)^2$
assuming a total cross section  of 40 mb.
Experimentally a maximum analyzing power of about 4.5\% at t$=-0.0037$ GeV/c was achieved 
which is shown in the lower part of Fig. \ref{dsan}. The data are from \cite{oka06}
taken with a 100 GeV/c proton beam at a polarized atomic hydrogen gas jet target.

For antiprotons the same analyzing power will result because at these energies the hadronic part is limited to the non spin-flip amplitude in the CNI region and the spin-flip Coulomb amplitude which changes the sign.
This was also experimentally verified in a measurement at FERMILAB with 185 GeV/c polarized antiprotons \cite{akc89} resulting in an analyzing power of $-4.6$ ($\pm$ 1.86) \%.
At 3.5 GeV/c momentum the assumptions for high energy may not be valid any more. The total elastic cross sections for $pp$ and $\bar{p}p$ at 3.5 GeV/c differ strongly which may indicate a stronger influence of additional amplitudes and could
result in a strong deviation from the high energy value of the analyzing power.
Also the total cross section of $\bar{p}p$ scattering is about a factor of 2 higher than compared to the 40 mb at high energy
which also indicates a shift of the CNI region to lower values.
Preliminary calculations of the analyzing power at lower beam momenta 
have been performed by Haidenbauer \cite{hai14b} in a one boson exchange model with the NN potential adjusted to experimental data of $\bar{p}p$ scattering \cite{hai14} which are available down to about 5 GeV/c.
The maximum in the analyzing power for the calculations for 5 GeV/c $\bar{p}$ momentum is at a t-value of about -0.0025 which is a bit lower than the  $t_p$ value at high energies but the maximum $A_{N}^{max}$ is still about 0.45 \%.
Therefore we expect for a 3.5 GeV/$c$ antiproton scattered on a proton an analyzing power comparable to
the high energy data of about 4.5\% at a laboratory scattering angle between 10 to 20 mrad.

\begin{figure}[hbt]
\begin{center}
\includegraphics[width=0.7\textwidth]{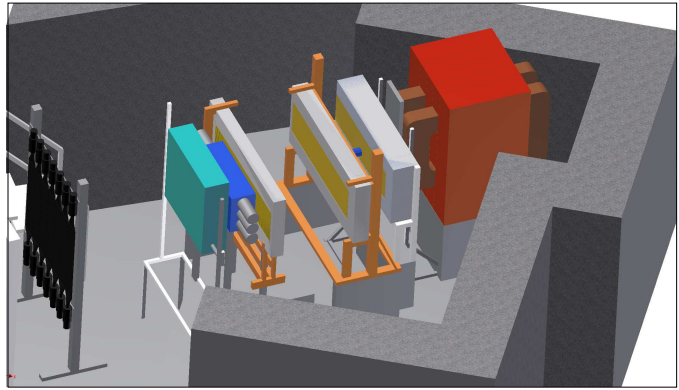} 
 \end{center}
\caption{T11 area with installed detector components. On the right side the last dipole of the beam line is seen from where the beam enters the detector system. 
\label{cernsetup}}
\end{figure}

In Fig. \ref{cernsetup} the setup for the proposed polarization study at the T11 beam line 
of the CERN PS complex is shown. 
 
 The T11 beam line delivers secondary particles produced by the 24 GeV/c momentum proton beam of the CERN/PS 
at a production angle of about 150 mrad with an acceptance of $\pm$ 3 mrad horizontally and $\pm$ 10 mrad vertically \cite{t11g}. The beam line can be adjusted to momenta of up to 3.5 GeV/c for positively and negatively charged particles. 
For positively charged particles of momenta of 3.5 GeV/c with open collimators the momentum resolution is $\pm$ 5 \% and up to $1 \cdot 10^6$ particles/spill are delivered with a setting for positively charged particles. The incident proton beam flux at this conditions is between $2 \cdot 10^{11}$ and $3 \cdot 10^{11}$ and the spill length is 400 ms.

From measurements at comparable conditions  \cite{eic72} (antiproton momentum 4 GeV/c, production angle 127 mrad)
a total flux of negatively charged particles of about $5 \cdot 10^5$/spill is expected which include about 4000 $\bar{p}$.

\begin{figure}[hbt]
\begin{center}
\includegraphics[width=\textwidth, clip=true]{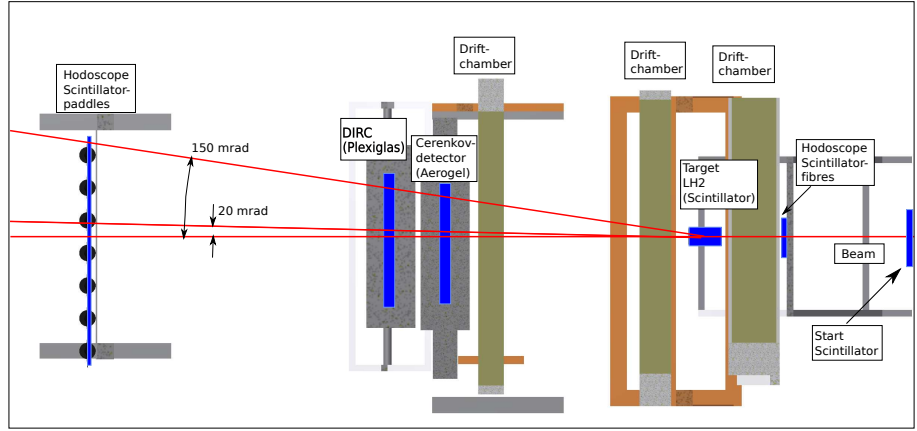} 
 \end{center}
\caption{Sectional drawing of the detector arrangement in the horizontal plane. The beam is entering from the right and hits start scintillator, beam hodoscope, first drift chamber, scattering target, drift chamber pair, Cherenkov detector, DIRC, and scintillator hodoscope.  A detailed description is given in the text.
The relevant scattering angle for the asymmetry measurement of 20 mrad is indicated by the red line. In total a scattering angular range of about 150 mrad is covered by the system. 
A similar angular range is covered in vertical direction.
\label{detsetup}}
\end{figure}

The detector arrangement for the measurement is shown in Fig. \ref{detsetup}.
It consists of scintillators for trigger signal generation and beam profile measurements, 
a drift chamber stack to measure the track of the produced antiproton, 
an analyzer target, a second drift chamber set to reconstruct the track of a scattered antiproton, a Cherenkov detector for pion discrimination, a DIRC for offline particle identification, and another scintillator hodoscope for trigger and $\bar{p}/\pi^-$ distinction by time of flight.
At the exit of the beam line tube which ends in front of the last beam line dipole a scintillation detector is mounted which will be used as a start detector for the timing and trigger generation.
A scintillation fibre hodoscope follows which will be used to determine the beam profile.
The drift chambers foreseen for these studies have been used in the \mbox{COSY-11} 
experiments at the cooler synchrotron COSY in J\"ulich \cite{brk96}.
The first chamber D1 has a hexagonal drift cell structure, optimized for low magnetic field sensitivity, with 3 straight and 4 inclined ($\pm$ 10 deg.) wire planes  \cite{smy05}.
The drift chamber set for the measurement of scattered antiprotons
include 4 (2) straight and 4 (4) inclined ($\pm$ 10 deg.) planes for D2 (D3).
 A track resolution in the order of 1 mrad was achieved with 1 GeV/c protons for D1 and the D2, D3 set.
 
\begin{figure}[hbt]
\includegraphics[width=0.5\textwidth, clip=true]{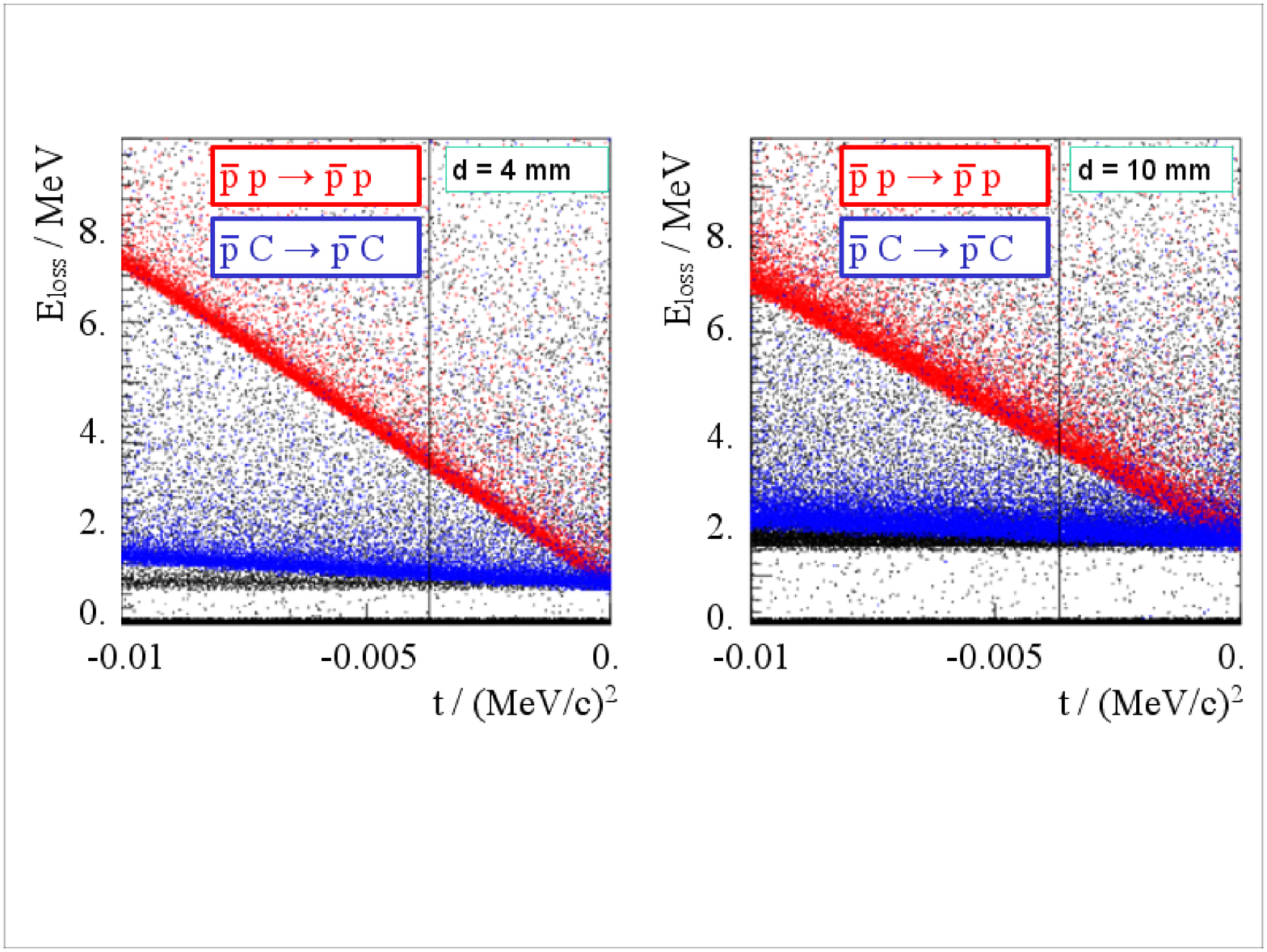} 
\hspace{1cm}
\begin{minipage}[b]{0.4\textwidth}
\caption{
Simulation of the energy loss in plastic scintillators of 4 mm thickness for 3.5 GeV/c antiprotons passing the 
scintillator without reaction (black dots), elastically scattered on a Carbon nucleus (blue dots) or 
elastically scattered on a proton (red dots) in the relevant t-range.
\vspace{1.5cm}
 }
\label{ppscat}

\end{minipage}

\end{figure}

As analyzer target a liquid hydrogen cell with a length of 15 cm will be used. The reconstructed tracks of primary and scattered antiproton allow to determine the reaction vertex with some uncertainty due to the limited track precision. For the analysis only the central part will be taken into account in order 
to discriminate reactions from the target windows.
As alternative an analyzer target consisting of several layers of 4 mm thick scintillators is foreseen.  
The carbon nuclei from the scintillator material will introduce a rather high background level 
but the scintillators can act as trigger for elastic $\bar{p} p$ scattering events in the relevant forward angular range.
An antiproton passing a scintillator will result in a minimum ionizing energy loss signal. In case of an elastic 
$\bar{p} p$ scattering process an additional energy loss from the scattered proton will be seen. In figure \ref{ppscat} 
the energy losses as a function of the t-value in plastic scintillators of 4 mm and 10 mm thickness are shown for 
Monte Carlo data of 3.5 GeV/c $\bar{p} p$ and $\bar{p} C$
scattering in addition to the expected energy loss of a minimum ionizing particle.
In the relevant t-range around $t = -0.0037$ GeV/$c^2$ a clear separation of $\bar{p} p$ events is visible but
quasi elastic processes like $\bar{p} C \rightarrow \bar{p} p X$ are not included in the MC data and will give rise to additional background signals. 
About 20 layers of scintillator modules are foreseen which result in a thickness comparable
to the liquid hydrogen target.

The whole system will be operated in air.  The straggling in the material between the first drift chamber and exit window of the beam tube does not influence the measurement.
The straggling in the material of the detection system is below the expected track resolution of the drift chambers.
For 3.5 GeV/c momentum antiprotons the radiation length up to the target
including D1 and the flight path of $\sim$ 0.5 m in air is about 30 $g/cm^2$ with a mean thickness of
0.13 $g/cm^2$ which introduces a straggling of about 0.2 mrad.
The tracking system for the
scattered antiprotons gives a straggling of below 0.8 mrad
and the hydrogen target adds another 0.7 mrad straggling.
With a 10 cm $CH$ target the straggling will be increased to 2 mrad,
which would still be low enough to measure the region of scattering angles around 20 mrad.

At the exit of the tracking system, an aerogel Cherenkov detector (n $\sim$ 1.03) to discriminate the expected
high pion background will be installed. The Cherenkov signals will be included in the trigger as veto.
When the event selection is reduced to single tracks at small scattering angles most background channels 
are suppressed and the momenta of the scattered particles are close to the momenta of the primary particles.
Antiprotons with 3.5 GeV/c momentum have a velocity of 
$\beta _{\bar{p}}$(3.5 GeV/c) = 0.966 , i.e. the threshold 
for Cerenkov light emission is at  n = 1.035.
For pions, which will be the main background source, the velocity is close 
to c ($\beta _{\pi}$(3.5 GeV/c) = 0.9992) with a threshold refractive index of 
n = 1.0008. Therefore a threshold Cerenkov detector with n $\sim$ 1.03
will drastically suppress the expected large pion background.

Behind the Cherenkov counter a DIRC with Plexiglas as radiator \cite{zin14} is mounted for offline particle identification.
The separation between antiprotons and pions at 3.5 GeV/c is expected to be 7.8 $\sigma$ deduced from recently performed
test measurement with a proton beam at COSY.
And finally a scintillator hodoscope
to trigger on single track events is positioned at a distance of about 6.8 m from the start scintillaor.
It consists of scintillator paddles with a width of 10 cm  readout on both ends by photomultipliers.

With the expected rate of up to $10^6$ particles/s at the T11 beam line 
a trigger signal has to be generated which selects antiproton scattering events while discriminating the background particles which are mostly pions.
An online track reconstruction to be used as event trigger would be very difficult with this event rates.
For a trigger signal we request a signal in the start scintillator and a signal in the scintillator hodoscope at the end and in addition a signal from the threshold Cherenkov detector as veto.
With this condition only charged particles which passes through the whole detector arrangement are detected and if the particle is a pion or a faster one it creates Cherenkov light with a high probability and blocks the trigger signal output.
From calculations and test measurements a pion discrimination by a factor 40 is expected which brings the event rate down to 25 000 events/s for the spill length of 400 ms which is sufficiently low that it can be stored by the data acquisition system without losses.
The extraction of elastic scattering events has to be done offline by reconstructing the tracks from the drift chambers and determine a possible kink in the track which gives the interaction point and the scattering angle.
We estimated a number of about 2.5$\cdot 10^{5}$ useful scattering event to be detected during the planned measurements.
Useful in the sense that they are in the t-range from $-0.002$ to $-0.007$ 
A rough estimate of the accuracy for an asymmetry measurement $\delta \epsilon$ with $N$ events is calculated by
$\delta \epsilon = \sqrt{(\frac{\delta \epsilon}{\delta L})^2 + (\frac{\delta \epsilon}{\delta R})^2}$
with $L=R=N/4$ for the events on the left (L) and right (R) side used to calculate $\epsilon = \frac{L - R}{L + R}$.
With 20\% polarization i.e. $\epsilon = P\cdot A_y =  0.009$ and 2.5$\cdot 10^{5}$ events an error   $\delta \epsilon$  of $\sim$ 30 \% results which is consistent to MC studies resulting in an asymmetry of $\epsilon = 0.012 \pm 24\%$
based on 2.5$\cdot 10^{5}$ events.

\section{Summary}

The polarization study of produced antiprotons will clarify the question if the antiproton production process creates some polarization. If so, it would be the ideal basis for the preparation a polarized antiproton beam.

The described detector setup was installed at the T11 beam line at CERN and a first measurement
was successfully completed in December 2014.
Preliminary analyses confirm the expected online pion discrimination and a clear antiproton separation in the offline reconstruction. The detailed data analysis will begin in 2015.


\end{document}